\documentstyle[aps,twocolumn,epsfig]{revtex}

\title{Counterfactual Statements and Weak Measurements: an Experimental Proposal}

\author{Klaus M\o lmer \\
{\small
  Institute of Physics and Astronomy, University of Aarhus}\\
{\small DK-8000 \AA rhus C}\\
{\small e-mail: moelmer@ifa.au.dk}}
\date{\today}

\begin{document}
\draft

\maketitle

\begin{abstract}
A recent analysis suggests that weak measurements can be
used to give observational meaning to counterfactual reasoning
in quantum physics. 
A weak measurement is predicted to assign a negative 
unit population to a specific state in an interferometric
Gedankenexperiment proposed by Hardy.
We propose an experimental implementation with trapped ions of the
Gedankenexperiment and of the weak measurement.
In our standard quantum mechanical analysis of the proposal no states
have negative  population, but we identify the  registration
of a negative population by particles being displaced on average in 
the direction opposite to a force acting upon them.
\end{abstract}

\bigskip

Pacs. 03.65 Bz, 87.70+c

\bigskip

In a Gedankenexperiment proposed by Hardy \cite{hardy} an electron
and a positron enter separate Mach-Zender-like interferometers, alligned
so that the particles leave the 
interferometers in definite  exit ports with unit probability.  
One of the internal arms
of the electron's interferometer is then made to overlap an arm of the 
positron's
interferometer, so that if both particles are present in those
respective arms they annihilate. 
In this setup, in addition to the possibility of loosing the 
particles, it is now possible to detect the particles in both
output arms of their interferometers.
A detection of the electron in the normally empty output port 
implies that the destructive interference of the two 
paths in the interferometer has been overruled.
Counterfactual reasoning suggests the explanation
that the positron must have been  
present in the interferometer arm that overlaps the electron's 
intereferometer and the 
electron must have  chosen the opposite arm to avoid annihilation.
If also the positron is detected in the normally empty output arm,
the same reasoning applies with the particles interchanged, thus we
conclude from such a joint measurement that the electron (and the
positron) took both one and the opposite path in the experiment:
Hardy's paradox \cite{notehardy}.

Hardy's paradox can be dealt with 
by a more careful analysis of what happens if a measurement device is inserted
in the electron interferometer so that we measure rather than infer the 
path taken by the particle. Quantum theory provides probabilistic
information about measurements, and a 
statement about a property of a system in the past,
such as 'the electron took the path without overlap with the positron
interferometer' is only correct and meaningful if it has been or could
as well have been measured. 
Inserting a detector inside the interferometer, however, changes 
the entire situation, and an analysis of the altered
system is not saying much about the original problem. 
But, how can we make quantitative statements about the 
past state of a system without such changes ?
A new ingredient in the theoretical analysis 
is a study of the role of a weak measurement
of the population of the different interference paths \cite{popescu}. 
Weak measurements are implemented by coupling to a quantum 'meter'
degree of freedom which changes by an amount proportional to the
value of the quantity measured. If the change of the meter 
is smaller than
its quantum mechanical uncertainty, a conclusive value cannot be
obtained from a single  measurement, but by repeating the experiment 
many times
it can be used to determine average properties of the system.
The advantage of the weak measurement for the present purpose
is that it does not disturb the state of the system significantly, we
can thus address the population of the interferometer arms
without destroying the interference.
Such measurements in coincidence with both particles leaving the 
interferometers in the normally empty output ports, are predicted 
in Ref.\cite{popescu} to yield an average
unit population for both paths where only one particle passes the 
annihilation region. The populations for all possible states add 
up to unity, and weak measurements should also yield an average 
minus unit population in the state where both particles avoid 
the annihilation region.

We suggest to implement the above scheme in the internal state dynamics
of two trapped ions. The experiment seems possible to do with present
technologies, and an analysis in terms of a well controlled
experimental system may be helpful to assess the 'meaning' of
negative population in weak measurements. As a supplement to 
Ref. \cite{popescu} which establishes a set of
self-consistent algebraic identities between operators and
weak values, we shall 
present a rather straightforward analysis of the quantum state of the
observed system, and the expectation values amenable to experimental
detection. Ref. \cite{popescu}  replaces Hardy's paradox with a
self-consistent, {\it albeit} strange, set of predictions. 
To the extent that we have learned to  understand interference of
wavefunction amplitudes,  the present  analysis removes the
strangeness of the predictions. 

Our first step is to replace the electron and the positron with two 
trapped ions, and to replace the motion in the interferometers
by the two-level dynamics of the internal electronic degrees of freedom
in the ions.
Both ions have three longlived internal states $|g\rangle,\ |e\rangle$ and 
$|f\rangle$, that we can couple coherently by means of laser fields,
and we assume transitions to a fluorescing level
used for cooling, for preparation of the initial state, and for final detection.
The ions are initially both in the internal state $|g\rangle$,
corresponding to the input ports to the interferometers.

We now describe the implementation of Hardy's interferometer:

\noindent
({\it i}) A resonant laser pulse on the $g-e$ transition in each ion 
implements the first beam splitters:
\begin{eqnarray}
|\Psi\rangle = |gg\rangle \rightarrow
\frac{1}{2}(|g\rangle+|e\rangle)(|g\rangle+|e\rangle).
\label{i}
\end{eqnarray}

\noindent
({\it ii}) A bichromatic laser pulse with  frequencies
$\omega_{\pm}=(E_f-E_e)/\hbar \pm (\nu-\delta)$
is applied to both ions, where $\nu$ is 
the center-of-mass vibrational frequency of the ions and $\delta$ 
is a detuning to avoid single sideband excitation of the ions. 
$\omega_++\omega_- = 2(E_f-E_e)/\hbar$, and as shown in
\cite{sorensen} the field will drive Rabi oscillations on the
resonant two-photon
transition $|ee\rangle \rightarrow |ff\rangle$, and 
after an appropriate interaction time
the ions are in the state
\begin{eqnarray}
|\Psi\rangle = \frac{1}{2}(|gg\rangle+|ge\rangle+|eg\rangle+|ff\rangle),
\label{ii}
\end{eqnarray}
in which the $|ee\rangle$ component has been 'annihilated'.

\noindent
({\it iii}) A second resonant laser pulse on the $g-e$
transition implements the second beam splitters. The state $|f\rangle$
is untouched, whereas $|g\rangle \rightarrow
\frac{1}{\sqrt{2}}(|g\rangle+|e\rangle),\ |e\rangle \rightarrow
\frac{1}{\sqrt{2}}(|e\rangle-|g\rangle)$, and the  resulting state vector
reads:
\begin{eqnarray}
|\Psi\rangle = \frac{1}{4}(2|ff\rangle + 3|ee\rangle + |ge\rangle +
|eg\rangle - |gg\rangle).
\label{iii}
\end{eqnarray}

As shown by (\ref{iii}), there is a
probability of $1/16$ to find the ions in the state  $|gg\rangle$,
and thus to get the same conflicting statements from counterfactual reasoning
as we obtained for the electron and positron interferometers.

We now turn  to the problem of measuring weakly the population of the
intermediate state $|gg\rangle$. Such 
weak measurements can, in general,
lead to surprising results far outside the range of eigenvalues of the
quantity measured \cite{weak,weakfarout} and in the present case ,
Ref. \cite{popescu} suggests
a minus unit population resulting from the measurement.
In order to determine the $|gg\rangle$ population 
after the steps ({\it i}) and ({\it ii}) described above
we couple 
the ions weakly to a meter. In the ion trap, the relative 
coordinate of the ions can be used as such a meter. The two ions will
have classical equilibrium positions at $\pm X_0$, 
situated symmetrically around the trap center. The motion around 
these positions $x_{1/2} = \pm X_0 + \delta x_{1/2}$
can be separated into a motion of the center-of-mass coordinate, and
a stretching of the distance between the particles, described by
$x_+ = \delta x_1 + \delta x_2$ and $x_-=\delta x_1 - \delta x_2$.
To an excellent approximation, the motion separates in harmonic
oscillator motion for $x_+$ and $x_-$, and the motional ground state
factorizes in a center-of-mass gaussian wave function of $x_+$ and a 
stretch-mode gaussian wave function of $x_-$.

\noindent
({\it iii'}) We apply a 
bichromatic laser field, which is off resonant with the excitation
from $|gg\rangle$ to the doubly excited state $|ff\rangle$ of the
ions so that the ions in $|gg\rangle$ are not excited, but they experience a 
lightshift (ac-stark shift). We assume that the laser 
field depends quadratically on position within the spatial region
occupied by the ions, so that the 
lightshift Hamiltonian can be written $\delta H = \kappa (x_1^2+x_2^2)
|gg\rangle\langle gg|$.  If the field is turned on adiabatically,
the $|gg\rangle$ component of the ions, will gradually feel
a change in the trapping potential, and if
$\kappa$ is positive  they will be shifted to a new
equilibrium position with the ions closer together, {\it i.e.}
with a negative mean value of $x_-$.
There will also be a scaling of the 
extent of the wave function of the ions, but
we shall ignore this in the following analysis.
Let $\phi(x_-)$ denote the relative spatial wave function of the
two ions before the action of the laser pulses. The beam splitter pulses
and the 'annihilation pulse' do not affect the spatial state,
but the 'meter' pulse causes a displacement of the
$|gg\rangle$ component by an amount $-a$, so that it is henceforth
described by $\phi(x_-+a)$. The center-of-mass wavefunction is not
affected and we shall omit it from our formal derivations. 
The state thus reads:
\begin{eqnarray}
|\Psi\rangle = 
\frac{1}{2}(|ge\rangle+|eg\rangle+|ff\rangle)\otimes\phi(x_-)+
\frac{1}{2}|gg\rangle\otimes\phi(x_-+a).
\label{iii'}
\end{eqnarray}

\noindent
({\it iv'}) We now have to determine the action of the second beam splitter
pulse (\ref{iii}), and this time we get:
\begin{eqnarray}
|\Psi\rangle = \frac{1}{4}(2(|ff\rangle+|ee\rangle-|gg\rangle)\otimes\phi(x_-) 
+\frac{1}{4}(|gg\rangle+|ge\rangle + |eg\rangle + |ee\rangle)\otimes\phi(x_-+a).
\label{iv'}
\end{eqnarray}

\noindent
({\it v'}) We detect the ions in the internal state $|gg\rangle$.
To avoid heating the atoms with the momentum recoil from this detection, 
one can check the absense of the ions
from the $|f\rangle$ and  $|e\rangle$ states by coupling those states to
a fluorescing level. If no fluorescence photons are emitted, the ions
must be in $|gg\rangle$.
Hereby we leave the {\it conditioned} relative motion of the ions in 
the state:
\begin{eqnarray}
\phi_c(x_-) = {\cal{N}}(\phi(x_-+a)-2\phi(x_-)),
\label{v'}
\end{eqnarray}
where $\cal{N}$ is a normalization constant.

The spatial wave function (\ref{v'}) is a superposition of the original 
state and
one shifted by $-a$.  We assume that the
light shift interaction is weak, and that $a$ is much smaller than the
width of the relative wave function of the ions. This means, that to
detect the $|gg\rangle$ population, one has to perform a large number of
measurements of the displacement $x_-$, and compute the average value.
Since the displacement is small
a first order Taylor expansion around $\phi(x_-)$ shows that
$\phi(x_-+a)+\phi(x_--a)=2\phi(x_-)$, and therefore
\begin{eqnarray}
\phi_c(x_-)=-\phi(x_--a).
\end{eqnarray}
The particles have moved further away by precisely the amount that
$|gg\rangle$ state ions would have moved closer together.
The Hamiltonian, responsible for this displacement is 
proportional to the projection operator on the state $|gg\rangle$ and we
can now see how the experiment can make a weak measurement
of a minus unit population in that state. One readily verifies, that if
the lightshift had been imposed on the $|ge\rangle$ or the $|eg\rangle$
state, in either case the ions would have been pulled closer together
by the average amount $a$, in correspondence with a unit population in both
states.

The experiment seems quite feasible. It is today possible to cool
ions to their motional ground state in an ion trap and to coherently
manipulate their internal and motional states \cite{cool2}. 
By use of Raman transitions, it is possible to
couple several stable states in the ions, and it is possible to
adjust the effective $k$-vector in the transition and the resulting
momentum transfer to the ions \cite{raman}.
All our interaction schemes are symmetric in the
internal states of the ions, and they can be implemented without
individual access to the ions. Both the beam splitter pulses and the
bichromatic pulses for simultaneous excitation of two ions have been
demonstrated in experiment \cite{sackett,kielpinski,rowe}, and the 
fluorescence technique 
provides perfect detection of the internal state. The meter for the 
$|gg\rangle$ detection is the relative coordinate between the ions. 
This relative coordinate is an eigenmode for the coupled ionic motion
in the trap, and it is well described by a harmonic oscillator.
The ground state width is on the order of a micron or less, 
and the experimental task is to measure a change in position which is even
smaller than that. In fact, in ion traps quantum state tomography has
been performed in order to determine the motional Wigner
function and the Fock state density matrix elements
\cite{tomography}. The resolution of these studies is indeed in the
range needed to distinguish displacements smaller than the ground state width.
So far, tomography was done on the motional state of a single trapped
ion, but there seems to be no principal problem in extending these
studies to the collective motion of two ions.
It is of course important to control the phases and the
timing of the detection, since 
the displaced state $\phi_c(x_-)$ is not an eigenstate 
of the Hamiltonian describing the system. One has to perform
the detection in a suitable interaction picture, as already done in
\cite{tomography}.

We return to the issue of precision {\it versus} disturbance
of the state vector:
The ground state of relative motion is a gaussian with width 
$\sigma$, $\phi(x_-)\propto
\exp(-x_-^2/4\sigma^2)$, where $x_-$ measures the separation of the ions
relative to the equilibrium value. 
In case of 'less weak'
measurements, {\it i.e.} with values of $a$ which are not
necessarily much smaller than $\sigma$, Eq.(\ref{v'}) gives: 
$\langle x_-\rangle =-a (1-2e^{-a^2/8\sigma^2})/(5-4e^{-a^2/8\sigma^2})$.
This expression yields $\langle x_-\rangle=+a$ for small $a$, as 
concluded above,
it changes sign to negative values when $a^2=(8\log 2) \sigma^2$,
indicating how 'strong' we can make the weak measurement,
and still obtain an appearant negative mean population of the intermediate
$|gg\rangle$-state. 

A simpler experiment can be imagined, where the 'meter'
is not the motional state of the ions, but the internal state of
a third ion in the trap. Such an ion could be prepared in a
superposition $(|g\rangle+|e\rangle)/\sqrt{2}$ of its two internal
states, and a Hamiltonian leading to the Toffoli- or $C^2-NOT$
gate of quantum computing \cite{toffoli} could be applied for a short time,
so that if the two ions are in the ground states, identified with the
logical value 1, the third ion undergoes a Rabi oscillation between its
states $|g\rangle$ and $|e\rangle$. Rather than the rotation of $\pi$
radians corresponding to a full $C^2-NOT$ operation, a small
positive rotation angle can be applied. In Ref.\cite{wang}, we
present an explicit Hamiltonian providing this rotation. 
The logic of the experiment is
now, that if the two ions are both in the state $|g\rangle$, the
$|e\rangle$ state population of the third ion increases by a definite
amount $\delta p$, and for exactly the same reason as above, one will
find the opposite result: the excited state population decreases by
$\delta p$.  Note that if
the third ion is chosen with a different transition frequency,
this experiment can also be made without individual access to the ions
in the trap.

Quantum theory is a cooking book telling us how to use wave functions
to make probabilistic predictions about outcomes of experiments.
This analysis can sometimes be supplemented by qualitative
explanations, where the skilled physicist succesfully mixes classical and
quantum mechanical reasoning to account for the results of
the quantitative theory or of an experiment.
Essentially, all the so-called paradoxes
in quantum physics stem from this approach, and they illustrate the
inadequacy of classical pictures of quantum processes rather than
any fundamental problems within quantum theory. A ''good paradox" may,
however, deepen our understanding of quantum behavior, and the actual
implementation of a physical experiment may be an important step
towards a real application of the physical effect behind the paradox.

We have proposed an experiment to address one such paradox.
It is noteworthy
that counterfactual reasoning which is not a consistent theoretical
framework, and which does not deal with observational facts, does
indeed bring an observational pecularity to our attention: the apparant
unit occupancy of two different states. This of course suggests that
other paradoxes raised by counterfactual reasoning be revisited with
weak measurement theory.
Apart from the natural interest in promoting what seems to be paradoxical
and unphysical effects to the status of observable facts,
this research may point to useful practical effects. 
In an optical experiment
\cite{weakfarout}, a large separation of 
two polarization components of a light field could be resolved easily 
on a weak signal. In principle the same information would be accessible
from stronger, less separated fields, but this would require
detectors operating in that intensity regime. Similar
trade-off issues may become relevant in the emerging field of quantum
computing \cite{issue}, where not only trapped ions, but also a number of
other physical systems offer the required access to single particle and
two-particle control, and where the above dynamics can therefore be
implemented.

This work was supported by the Information Society Technologies
programme IST-1999-11053, action line 6-2-1 EQUIP.
Discussions with Sandu Popescu and with Anders S\o rensen are gratefully
acknowledged.


\begin{thebibliography}{99}
\bibitem{notehardy} The analysis in Ref.[1] proceeds with a discussion
of paradoxical effects associated with the treatment of the detection
processes in different Lorentz frames. Our work, however, deals only
with the situation described for a single laboratory frame.
\end{thebibliography}
\end{document}